\documentclass[twocolumn,aps,prb,showpacs,superscriptaddress,preprintnumbers]{revtex4-2}
\usepackage{amsfonts}
\usepackage{amsmath}
\usepackage{amssymb}
\usepackage{multirow}
\usepackage{graphicx}% Include figure files
\usepackage{dcolumn}% Align table columns on decimal point
\usepackage{bm}% bold math
\usepackage{xcolor}
\usepackage{soul}

\begin{document}

\title{A composite ansatz for the calculation of dynamical structure factor}

\author{Yupei Zhang}
\affiliation{HEDPS, Center for Applied Physics and Technology, and School of Physics, Peking University, Beijing 100871, China}

\author{Chongjie Mo}
\email{cjmo@csrc.ac.cn}
\affiliation{Beijing Computational Science Research Center, Beijing 100193, China}
%\affiliation{Tianfu Innovation Energy Establishment, Chengdu 610210, China}

\author{Ping Zhang}
%\email{zhang_ping@iapcm.ac.cn}
%\affiliation{Tianfu Innovation Energy Establishment, Chengdu 610210, China}
\affiliation{Institute of Applied Physics and Computational Mathematics, Beijing 100088, China}
\affiliation{HEDPS, Center for Applied Physics and Technology, and College of Engineering, Peking University, Beijing 100871, China}

\author{Wei Kang}
\email{weikang@pku.edu.cn}
\affiliation{HEDPS, Center for Applied Physics and Technology, and College of Engineering, Peking University, Beijing 100871, China}
%\affiliation{Tianfu Innovation Energy Establishment, Chengdu 610210, China}

\date{\today}

\begin{abstract}
We propose an ansatz without adjustable parameters for the calculation of dynamical structure factor.
The ansatz combines  quasi-particle Green's function, especially the contribution from the renormalization factor, and  the exchange-correlation kernel from time-dependent density functional theory together, verified for typical metals and semiconductors from plasmon excitation regime to Compton scattering regime. 
It has the capability to reconcile both small-angle and large-angle x-ray scattering (IXS) signals with much improved accuracy, which can be used, as the theoretical base model, in inversely inferring electronic structures of condensed matter from IXS experimental signals directly.
It may also used to diagnose thermal parameters, such as temperature and density, of dense plasmas in x-ray Thomson scattering experiments.

\end{abstract}

\maketitle
% Copyright (c) 2014,2016 Casper Ti. Vector
% Public domain.
\section{Introduction}

Dynamical structure factor (DSF) $S({\bf q},\omega)$ is a straight-forward description for correlated motion of electrons in solids \cite{Onida2002,Schulke2007,Panholzer2018} 
and dense plasmas \cite{Glenzer2003,Glenzer2009,Frydrych2020,Astapenko2021}, and whereby of persisting interests in a variety of research branches
\cite{Li2021,Shen2017High,Lee2019Oxygen,Hagiya2020,Hill1996,Reed2010,Bradley2010,Wang2020}.
Since it can be directly detected in inelastic x-ray scattering (IXS) experiments \cite{Schulke2007,Weissker2006,Eisenberger1975,Schulke1984,Schulke1986,Sturm1992} or electron energy loss experiments \cite{egerton2011electron,RevModPhys.82.209} with a resolution on both transferred momentum $\bf q$ and frequency $\omega$, 
DSF has been long considered as a prospective tool to infer electronic structures and collective excitations in a many-electron system.
In condensed materials, a number of properties associated with electronic structures and excitations, e.g., the plasmon excitation \cite{Schulke1984,Hambach2008,Cazzaniga2011,Panholzer2018}, the double-plasmon excitation \cite{Spence1971,Schattschneider1987,Sternemann2005,Huotari2008,Panholzer2018}, the near-edge structure \cite{Sternemann2007,Sahle2013,Fister2009,Fister2011}, the single-particle excitation continuum \cite{Huotari2011}, and even part of the excitonic effect \cite{abbamonte2008dynamical,Sundermann2018}, are considered to present in the DSF with detectable features. 
In dense plasmas, DSF, usually probed in x-ray Thomson scattering experiments, is considered as an important, and sometimes irreplaceable diagnostic tool to probe thermal properties, such as temperature and density, in the internal regime of dense plasmas \cite{Glenzer2009,Sperling2015,Mo2018,Mo2020,Falk2018,Dornheim2022}.

It is a common practice to calculate DSF in the framework of time-dependent density functional theory (TDDFT) \cite{Runge1984,Petersilka1996}, which avoids the cumbersome calculation of Bethe-Salpeter equation \cite{Onida2002} of the many-body perturbation theory (MBPT) approach, owing to the less important contribution of long-range exchange and correlation effect in the DSF  \cite{Onida2002}. 
It was shown \cite{Weissker2006,Weissker2010,Huotari2011,Cazzaniga2011,Seidu2018} as a rule of the thumb that for $|{\bf q}|$ smaller than twice of the plasmon cutoff wave vector $q_c$, the method reproduced the experimental DSF well with life-time broadening effects included semi-empirically. 
Beyond that, theoretical results significantly differ from experimental measurements.
It thus has been longed for having a theoretical approach that is able to reconcile DSF to much larger $|{\bf q}|$, so that large-angle x-ray scattering signal may also be well interpreted.
If so, the capability of various light sources, including synchrotron light sources \cite{Lee2019Oxygen,Hill1996,Reed2010,Bradley2010,Wang2020,Weissker2006,Schulke1984,Hambach2008,Sternemann2005,Sahle2013,Fister2011,abbamonte2008dynamical} and the free-electron laser light sources \cite{Ding2009,Frydrych2020} would be much strengthened.

In this work, we propose an ansatz with no adjustable parameter to reconcile DSF measurements, verified for valence electrons of metals and semiconductors (including lithium, sodium, aluminum, and silicon)  from plasmon excitation regime to Compton scattering regime. 
The ansatz contains two major ingredients: the contributions of quasi-particle Green's function, including those from quasi-particle energy-level shifts, life-time broadening, and renormalization factors \cite{Hedin1965,Hybertsen1986,Northrup1987}; together with the corrections from the TDDFT kernel, which contributes the effective local field effect \cite{Weissker2006,Weissker2010,Huotari2011,Cazzaniga2011}.
We shall show that the contributions of the quasi-particle Green's function and of the TDDFT kernel, although originated from distinct theoretical frameworks, (i.e., from MBPT and TDDFT respectively,) can be combined together to arrive at a simple but effective formula.

The rest of the article is organized as follows. In Sec. II, the composite ansatz and corresponding formulas are presented. 
In Sec. III, we provided numerical details for carrying out the calculation with the new ansatz. 
In Sec. IV, The performance of the ansatz in predicting DSF from small angle to large angle scattering is illustrated with lithium and silicon. Also provided are physical discussions on where the improvements come from. 
In Sec. V, we conclude the article with a concise summary.

\section{Formulas of the Composite ansatz}

Formally, $S({\bf q},\omega)$ can be derived from the response function $\chi({\bf q},{\bf q}',\omega)$ through the fluctuation-dissipation theorem \cite{PinesB1966}, i.e., $S({\bf q},\omega)=-\Im [\chi({\bf q},{\bf q}',\omega)]|_{\bf{q}'={\bf q}}/(\pi n)$, where $n$ is the electronic density.
The screening effect is accounted for by a Dyson-like equation \cite{Petersilka1996}
\begin{eqnarray}\label{dyson}
\chi({\bf q},{\bf q}',\omega)&= &\chi_{0}({\bf q},{\bf q}',\omega)+\int d{\bf q}_{1} d{\bf q}_{2}\, \chi_{0}({\bf q},{\bf q}_{1},\omega) \nonumber \\
&\times& K({\bf q}_{1},{\bf q}_{2},\omega)\chi({\bf q}_{2},{\bf q}',\omega),
\end{eqnarray}
where $\chi_{0}({\bf q},{\bf q}',\omega)$ is the bare response function, and $K({\bf q},{\bf q}',\omega)$ is the many-body interaction kernel.

In the ansatz, $\chi_0$ and $K$ are treated separately.
The kernel $K({\bf q},{\bf q}',\omega)$ is approximated as the sum of Coulomb contribution $v_c(q)=4\pi/|{\bf q}|^2$ and an effective contribution containing exchange and correlation effects $f_{xc}({\bf q},{\bf q}',\omega)$ from TDDFT, i.e.,
\begin{equation}\label{Keq}
K({\bf q},{\bf q}',\omega)=v_c(q)\delta({\bf q}-{\bf q}')+f_{xc}({\bf q},{\bf q}',\omega).
\end{equation} 
Eqs.~(\ref{dyson}) and (\ref{Keq}) will be reduced to the random phase approximation of $\chi$ when $f_{xc}$ is set to be zero.

The bare response function $\chi_0({\bf q},{\bf q}',\omega)$ is calculated as the Fourier transform of $\chi_0({\bf r},t;{\bf r}',t')$, where $\bf r$ and ${\bf r}'$ are coordinates in the three-dimensional space, and 
\begin{equation}\label{chi0_eq}
\chi_0({\bf r},t;{\bf r}',t')=-iG({\bf r},t;{\bf r}',t')G({\bf r}',t';{\bf r},t)/\zeta.
\end{equation}
Here, $G({\bf r},t;{\bf r}',t')$ is the screened Green's function,  and $\zeta$ is an amplitude scaling constant to  numerically guarantee the sum rule $\int_{-\infty}^{\infty} \omega S({\bf q},\omega) d\omega = |{\bf q}|^2/2 $ (atomic units) \cite{mahan2000many}. 
With the help of a set of quasi-particle wave functions, denoted as $\phi_j({\bf r})$ with $j$  the index of wave functions, $\chi_{0}({\bf r},{\bf r}';\omega)$ can be recast as 
\begin{eqnarray}\label{chi}
\chi_{0}({\bf r},{\bf r}';\omega)&=
&\sum_{j\neq k}(f_{j}-f_{k})\frac{|Z_{k}||Z_{j}|}{\zeta}\nonumber\\
&\times&\frac{\phi_{k}({\bf r})\phi_{j}^{*}
	({\bf r})\phi_{j}({\bf r}')\phi_{k}^{*}({\bf r}')}
{\omega-(\epsilon^{QP}_{k}-\epsilon^{QP}_{j})},
\end{eqnarray}
where $\epsilon^{QP}_j$ and $Z_j$ are quasi-particle energy and renormalization factor associated with the $j$-th quasi-particle state, and $f_j$ is the effective occupation number.
$\zeta$ is then determined as the product of the renormalization factors of the highest quasi-holes (denoted as $Z^{-}$) and of the lowest quasi-electrons (denoted as $Z^{+}$) with respect to the Fermi level, i.e., $\zeta=|Z^{+}Z^{-}|$.
It is crucial in the ansatz to set $\zeta$ in this particular form, which not only recovers the sum rule, but also accounts for the significant improvements in the DSF spectra at high transferred momenta.

\section{Computational Details}

In practice, the quasi-particle properties are calculated using the GW method \cite{Hedin1965,Hybertsen1986}. 
In particular, we follow Ref.~\cite{Hybertsen1986} for the calculation of $\epsilon^{QP}_j$, $Z_j$ and $\phi_j({\bf r})$, which means that $\phi_j({\bf r})$ is approximated by the corresponding Kohn-Sham wave function $\phi^{KS}_j({\bf r})$ , $\epsilon^{QP}_{j} = \epsilon^{KS}_{j}+Z_{j} \langle\phi_{j}|\Sigma(\epsilon^{KS}_{j})-V^{KS}_{xc}|\phi_{j}\rangle$, and $Z_{j}^{-1}=1-\left.\frac{\partial\Sigma_{j}(\omega)}{\partial\omega}\right|_{\omega=\epsilon^{KS}_{j}}$.
Here, $\Sigma$ is the self-energy operator, $\epsilon^{KS}_j$ is the Kohn-Sham eigen-energy associated with $\phi^{KS}_j$, and $V^{KS}_{xc}$ is the exchange-correlation potential. 
Note that now both $Z_j$ and $\epsilon^{QP}_j$ are complex numbers, and the imaginary parts of $\epsilon^{QP}_j$ for quasi-holes and quasi-electrons have opposite signs. 
So, Eq.~(\ref{chi}) naturally contains a damping coefficient $\eta=\Im(\epsilon_k^{QP}-\epsilon_j^{QP})=|\Im\Sigma_{k}|+|\Im\Sigma_{j}|$ \cite{Huotari2011,Cazzaniga2011,Weissker2006,Weissker2010}.

Here, we take lithium (Li) and silicon (Si) crystalline structures as demonstrating examples to display the performance of the new method, where Li and is a typical metal and Si is considered as a prototype of semiconductors.
%We first display the performance of this method, taking lithium (Li) and silicon (Si) crystalline structures as demonstrating examples, where Li is a typical metal and Si is considered as a prototype of semiconductors.
In the calculations, the lattice constant is set to be 6.6 bohr for Li and 10.26 bohr for Si, taken from measured values at room temperature \cite{Frank1996,touloukian1975-1,li1980refractive,Ashcroft1976}. 
Kohn-Sham eigen-energies $\epsilon_{j}^{KS}$ and corresponding wave functions $\phi_{j}^{KS}$ are obtained from a ground-state density functional theory (DFT) calculation via the \texttt{Quantum Espresso} package \cite{Giannozzi_2009,Giannozzi_2017}, using  norm-conserving pseudopotentials with only valence electrons included, i.e., one $2s$ electron for Li and four $3s3p$ electrons for Si.
The cutoff radii of the pseudopotentials are 2.97 bohr for Li and 1.8 bohr for Si, which yields convergent results at a plane wave cutoff energy of 80 Ry for both Li and Si, as have been carefully examined.
In order to have a close comparison with preceding results \cite{Weissker2006,Weissker2010}, a local density approximation version of exchange-correlation functional is used.
To match the transferred momentum ${\bf q}$ in experimental measurements, a much refined Monkhorst-Pack $\bf k$-point mesh \cite{Monkhorst1976} is adopted.
It varies from 8$\times$8$\times$8 to 14$\times$14$\times$14 so that one can find a $\bf q$ in the calculation which is close to the experimental transferred momentum with a difference less than 3\%.

\begin{figure}
	\centering
	\includegraphics[width=0.49\textwidth]{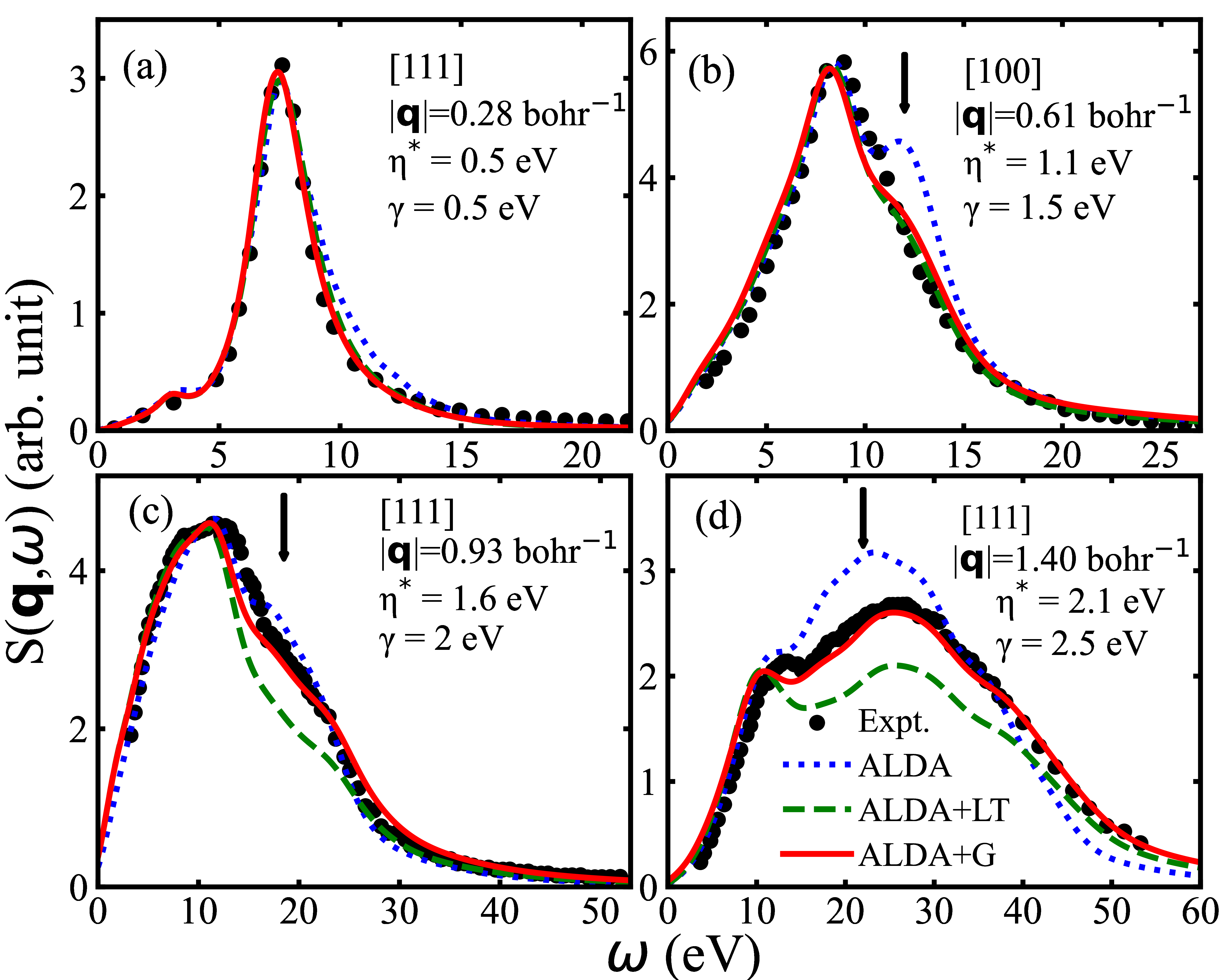}
	\caption{Calculated DSF of Li together with IXS spectra measured by Sch$\rm\ddot{u}$lke {\it et al.} \cite{Schulke1984,Schulke1986} at selected $|\bf q|$ along [100] and [111] directions, where $\eta^*$ is the broadening factor used in the TDDFT calculation with the ALDA kernel, and $\gamma$ is the Gaussian broadening factor to account for the finite resolution in $\omega$ caused by discrete sampling of Brillouin zone.}
	\label{fig:fig1}
\end{figure}

All DSF calculations, including calculations of quasi-particle properties using the GW method, are carried out using the \texttt{Yambo} code \cite{Yambo2009,Sangalli_2019} with necessary revisions.
In the GW calculations, the screening cutoff is 10 Ry, and the self-energy cutoff is 15 Ry with 240 bands included.
When $\chi$ is calculated via Eqs.~(\ref{dyson}) and (\ref{Keq}), the adiabatic local density approximation (ALDA) \cite{Gross1985} for $f_{xc}$ is adopted.
A more sophisticated form of $f_{xc}$ is also possible, but turns out having a small effect on metallic and typical semiconductor materials.
The DSF spectra presented are further smoothed with a Gaussian smearing $\gamma$ varying from 0.5 to 2.5 eV following Ref.~\cite{Mo2018} to account for the finite resolution in $\omega$ caused by the discrete sampling of Brillouin zone.
The smearing increases with $|{\bf q}|$, and is explicitly indicated in each calculation.

\section{Results and Discussions}
\subsection{DSF of Li}

Figure.~\ref{fig:fig1} shows the calculated DSF of Li, along the [100] and [111] directions.
The transferred momenta vary from 0.28 to 1.40 bohr$^{-1}$.
In the figure, the results of the new method are displayed as ``ALDA+G’’ and red solid  curves, meaning that all features of the Green’s function are included in conjunction with an ALDA $f_{xc}$ in Eq.~(\ref{Keq}).
The results of the original TDDFT method are displayed as ``ALDA’’ and blue dotted curves, and those calculated with ALDA $f_{xc}$ and life-time corrections, as proposed in Ref.~\cite{Huotari2011,Cazzaniga2011}, are denoted as ``ALDA+LT’’ together with green dashed  curves.
For comparison purposes, IXS measurements \cite{Schulke2007,Schulke1984,Schulke1986} are also presented in the figure, displayed as solid scattering dots.

Figure.~\ref{fig:fig1}(a) shows the capability of the new method to capture the feature of plasmon peak at small but finite $|{\bf q}|$, which is located around 7 eV. 
It shows that the new calculation closely follows the experimental data.
In this regime, it has been also known in practice that the TDDFT method with the ALDA kernel reproduces the experimental spectrum well for metals, as displayed by the dotted curve.

Figure.~\ref{fig:fig1}(b) illustrates the necessity of life-time broadening in metallic systems, where $|{\bf q}|$ is between $q_c$ and $2q_c$.  
Note that $q_c$ for Li is about 0.46 bohr$^{-1}$\cite{Schulke2007}. 
An important feature of the spectrum is the narrow shoulder at about 12 eV, as indicated by the arrow.
Without life-time effects, the ALDA calculation presents a separate small peak, which is clearly different from the narrow-shoulder feature of the experimental spectrum (displayed as scattering points.)
This deviation can be remedied by including the life-time broadening effect, i.e., the imaginary part of quasi-particle energies, as has been pointed out in the work of Weissker {\it et al.} \cite{Weissker2006,Weissker2010}.
The figure shows that both the new ALDA+G method and ALDA+LT method, which take the life-time broadening into consideration, indeed smooth out the peak and give a much closer agreement to the experimental spectrum.
It also shows that other quasi-particle features in the new method, such as the quasi-particle energy shift and the renormalization factor, do not contribute significant corrections to the spectrum at this $|{\bf q}|$.

\begin{figure}
	\centering
	\includegraphics[width=0.5\textwidth]{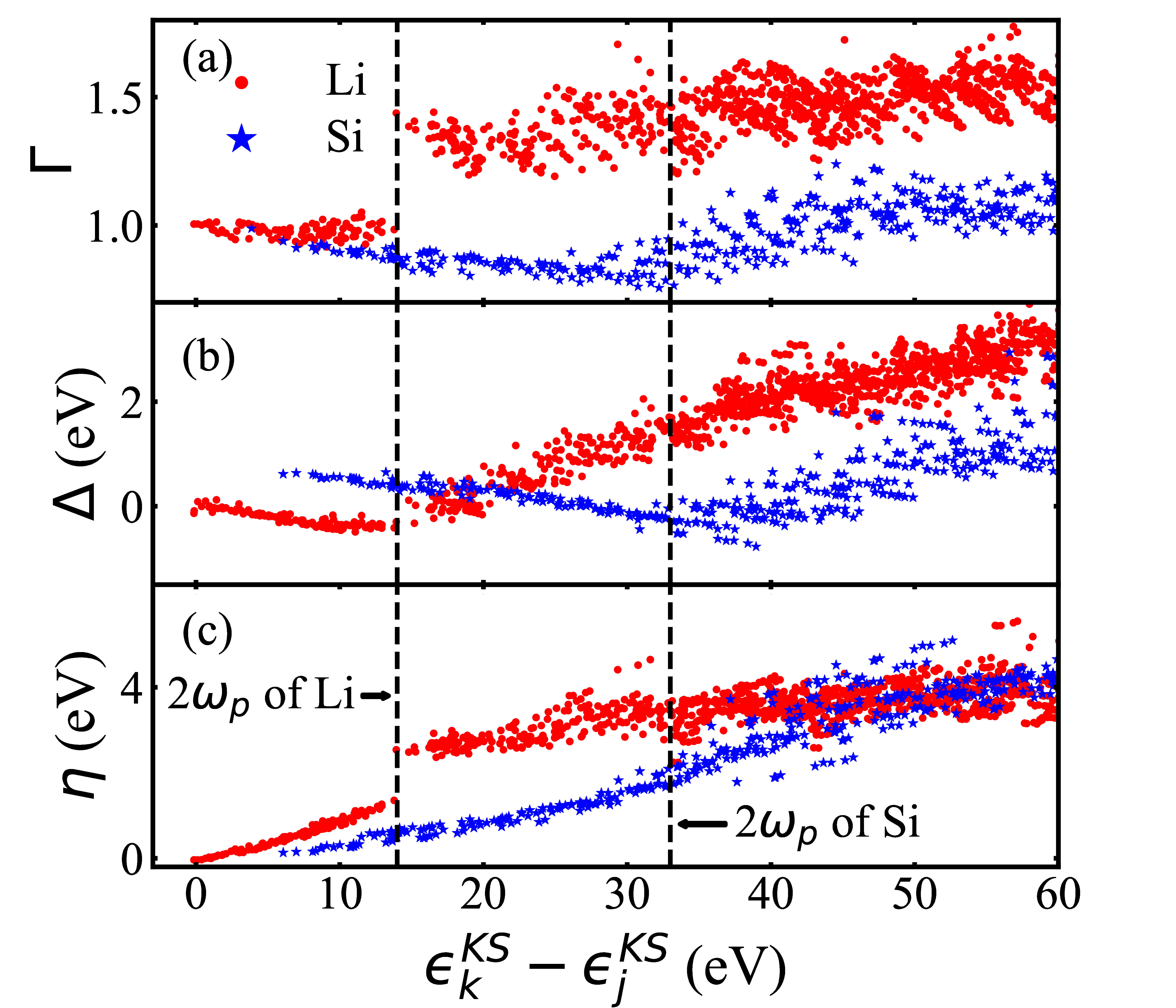}
	\caption{Rescaling coefficient, quasi-particle energy shift, and life-time broadening of Li and Si, calculated for electron-hole pairs of the GW method. (a) The rescaling coefficient $\Gamma$, defined as $\Gamma=|Z_k||Z_j|/\zeta$. (b) The quasi-particle energy shift $\Delta$, defined as $\Delta=\Re(\epsilon_k^{QP}-\epsilon_j^{QP})-(\epsilon_k^{KS}-\epsilon_j^{KS})$. (c) The life-time broadening $\eta$, defined as $\eta=\Im(\epsilon_k^{QP}-\epsilon_j^{QP})$. Note that forbidden transitions are not presented in the figure.}
	\label{fig:fig2}
\end{figure}

However, when $|{\bf q}|$ further increases to far above $2q_c$, the life-time broadening becomes insufficient to account for experimentally measured features.
As shown in Figs.~\ref{fig:fig1}(c) and (d), the ALDA+LT results generally underestimate the strength of the spectra at high energy.
But the detailed features of the calculated spectrum, i.e., the fluctuations and ripples, are quite similar to those of the measured spectrum if the strength underestimation were not considered, see e.g., the second peak indicated by the arrows in Fig.~\ref{fig:fig1}(c) and (d).  
It suggests that one may lift the ALDA+LT spectrum strength at high energy by introducing a scaling parameter in order to reproduce the experimental results.
It turns out that the renormalization factor can serve the purpose in a simple way.
It enters the ansatz as part of the screened Green's function $G$, as displayed in Eqs.~(\ref{chi0_eq}) and (\ref{chi}), and the scaling effect is presented by the combination of $|Z_k||Z_j|/\zeta$ in Eq.~(\ref{chi}), which will be called rescaling coefficient $\Gamma$ hereafter.
Fig.~\ref{fig:fig1}(c) and (d) show that when $\Gamma$ is included according to Eq.~(\ref{chi}), the theoretical spectra are substantially improved and in close agreement with experimental results up to $|{\bf q}|\approx$ 1.40 bohr$^{-1}$, the highest transferred momentum measured for Li so far.
DSF of other transferred momenta and directions are provided in the Supplemental Materials \footnote{See Supplemental Materials for other transferred momenta and directions of Li in Sec. I.}, showing a similar trend as that displayed in Fig.~\ref{fig:fig1}.

\subsection{Inheritance relation between the ansatz and previous methods}
From the theoretical results in Fig.~\ref{fig:fig1}, one may have noticed that there is a hierarchy of approximations for the calculation of DSF with increasing sophistication and application range of $|{\bf q}|$.
It starts from the TDDFT method with an ALDA kernel, and then is improved by including life-time broadening effects.
A further improvement is achieved by taking into consideration the rescaling effect of the renormalization factor, through the inclusion of the full screened Green's function.
With the proposed ansatz, this hierarchy can be understood as a series of simplifications in the calculation of $\chi_0$ in Eq.~(\ref{chi}).
The TDDFT method amounts to set the rescaling coefficient $\Gamma$ to be 1 and substitute the quasi-particle energies $\epsilon^{QP}_{j,k}$ by the Kohn-Sham eigen-energies $\epsilon^{KS}_{j,k}$.
Note that, in practice, the TDDFT calculation is usually carried out with a small constant life-time broadening \cite{Timrov2013}.
Further including the life-time broadening effect leads to the TDDFT method with life-time correction proposed by Weissker {\it et al.} \cite{Weissker2006,Weissker2010}.
When all the features of the screened Green's function are taken into consideration, the complete formula of the ansatz is recovered.

In order to understand how the hierarchy of approximations work, we plot the quasi-particle properties, i.e., the rescaling coefficient $\Gamma$, the quasi-particle energy shift $\Delta=\Re(\epsilon_k^{QP}-\epsilon_j^{QP})-(\epsilon_k^{KS}-\epsilon_j^{KS})$, and the life-time broadening $\eta=\Im(\epsilon_k^{QP}-\epsilon_j^{QP})$ of electron-hole pairs in Fig.~\ref{fig:fig2}, with respect to the transition energy $(\epsilon_k^{KS}-\epsilon_j^{KS})$ calculated using the DFT method.
It shows that there is an abrupt change taking place at a transition energy around $2\omega_p$ for all the three properties.
Note that $\omega_p$ for Li is 7.12 eV \cite{kittel_book}.
Below the energy, which is about 15 eV for Li, $\Gamma$ is roughly a constant about 1, the quasi-particle energy shift is close to zero, and the life-time broadening increases monotonically from zero to about 1.4 eV.
Above the  energy, the rescaling coefficient is rapidly lifted to a value between 1.2 and 1.7, the quasi-particle energy shift starts to increase linearly, and the life-time broadening jumps to a value in the range from 2.3 to 4.7 eV.

These features help to determine the validity regime of the series of approximations.
For example, for the metallic Li, the main feature of the DSF spectrum at small $|{\bf q|}$ is a plasmon peak centered at around 7 eV with a full width of half maximum about 3 eV, as displayed in Fig.~\ref{fig:fig1}(a). 
In this limited energy range, the rescaling coefficient is around 1, the quasi-particle energy shift is close to zero, and the life-time broadening is between 0.45 to 0.75 eV.
When this energy-dependent life-time broadening is approximated by a small constant broadening factor about 0.6 eV, the calculation is reduced to a TDDFT calculation in common practice.

\begin{figure}
	\centering
	\includegraphics[width=0.49\textwidth]{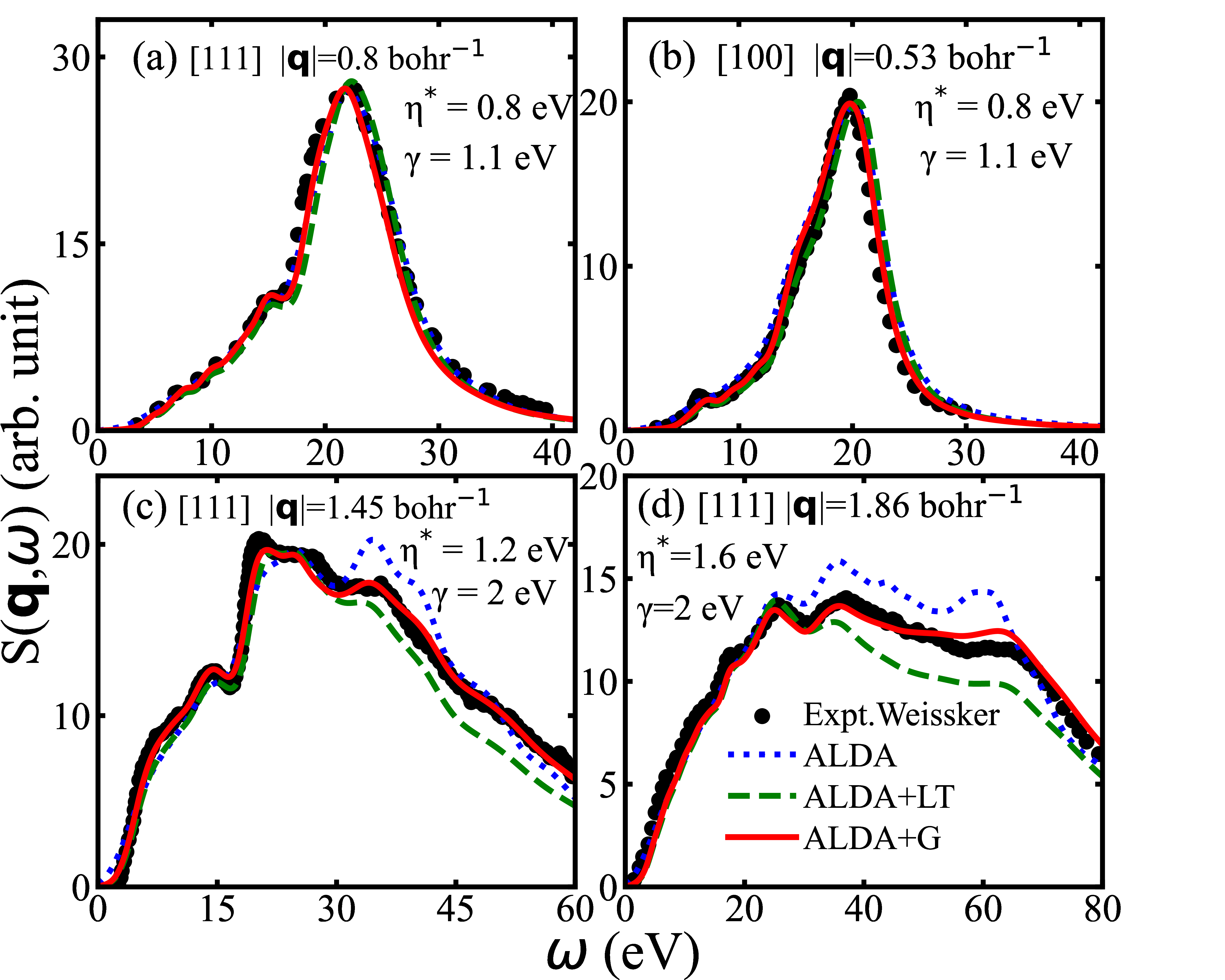}
	\caption{ Calculated DSF together with IXS spectra measured by  Weissker {\it et al.} \cite{Weissker2006,Weissker2010} at selected $|\bf q|$ from 0.53 bohr$^{-1}$ to 1.86 bohr$^{-1}$ along [100] and [111] directions. The ALDA results are calculated with a constant broadening $\eta^*$ as indicated. Additional Gaussian broadening of width  $\gamma$ indicated is carried out  at the end of each calculation to account for finite resolution of $\omega$.}
	\label{fig:fig4}
\end{figure}

However, when the spectrum covers a larger energy range, as it usually does when $|{\bf q|}$ increases, a constant life-time broadening is not enough to account for the broadening of the spectrum for the entire energy range.
In particular, when $|{\bf q}|$ increases to some magnitude such that the main feature of the spectrum is located in an energy range from 0 to about $2\omega_p$, i.e., about 15 eV for Li, it would be necessary to take the full energy-dependent life-time broadening into consideration to arrive at a good approximation, i.e., the life-time corrected TDDFT method proposed by Weissker, \textit{et al.} \cite{Weissker2006,Weissker2010,Huotari2011,Cazzaniga2011,Seidu2018} is recovered, since in this energy range the rescaling coefficient is still around 1 and the quasi-particle energy shift is close to zero.
This is exactly what Fig.~\ref{fig:fig1}(b) and Fig.~\ref{fig:fig2} have been illustrated, where the rescaling coefficient is about 1, and the quasi-particle energy shift is less than 0.5 eV.

Further increasing $|{\bf q}|$ makes the spectrum range go beyond the $2\omega_p$ threshold, as displayed in Fig.~\ref{fig:fig1}(c) and (d), it is then necessary to take account of the full feature of the screened Green's function. 
Fig.~\ref{fig:fig2} shows that in addition to the significant change of the quasi-particle energy shift and life-time broadening, the rescaling coefficient also experiences an abrupt increase from 1 to a value between 1.2 and 1.7 when crossing the $2\omega_p$ threshold, as the result of a sharp increase of the renormalization factor $|Z|$ of quasi-electrons from 0.6 to a value between 0.8 and 1.0 \footnote{The modulus of renormalization factor $Z$ and other quasi-particle properties can be found in Supplemental Materials}.
Under this condition, the complete form of the ansatz in Eq.~(\ref{chi}), which includes the variation of renormalization factor, becomes indispensable for a satisfactory prediction of DSF spectra.
  
\subsection{Effects of $\Gamma$, $\Delta$, and $\eta$ in large angle scattering}

From Fig.~\ref{fig:fig2}, one can also see how the strength of the DSF affected by the three quantities, i.e., the rescaling coefficient $\Gamma$, the quasi-particle energy shift $\Delta$, and the life-time broadening $\eta$. 
Since the major corrections are prominent for large transfer momentum $\bf q$, where $\chi$ can be well approximated by $\chi_{0}$ according to Eq.~\ref{Keq}, as can be seen if one notices that the kernel $K$ decays to zero as $q^{-2}$ at large $|{\bf q}|$ \cite{Onida2002} in Eq.~\ref{Keq}. 
Using the spectrum of Li in Fig.~\ref{fig:fig1}(d) as an example, ALDA results which do not have any of the three corrections, overestimate the spectrum strength between $\sim$ 15 eV to $\sim$ 30 eV, when compared with experiments. 
The ALDA+LT results, which includes the corrections of both quasi-particle energy shift $\Delta$ and life-time broadening $\eta$, but without the correction of rescaling coefficient $\Gamma$, show that the two quantities provide a smooth effect to the DSF strength through the expression of $\chi_{0}$ in Eq. (4), for energy greater than $\sim$ 12 eV.  

On one hand, the monotonic increasing of $\Delta$, which takes place from $\sim$ 12 eV for Li, as displayed in Fig.~\ref{fig:fig2}(b), decreases the joint density of states, which is proportional to the strength of DSF. 
On the other hand, $\eta$ provides an additional increasing of broadening effect, which further decreases the strength of DSF. 
The overall effects of these two quantities, however, give rise to an overshoot, resulting in an underestimated DSF strength compared to the experiments. 
Fig.~\ref{fig:fig2}(a) shows that the rescaling coefficient $\Gamma$ increases from about 1 to 1.5, which gives a boost to the strength of DSF to recover the experimental spectrum for energy greater than 12 eV.
 
 \subsection{DSF of Si}
The new method also applies to the DSF calculation of typical semiconductors.
Its capability to provide much improved predictions is displayed in Fig.~\ref{fig:fig4} using Si as an illustrating example.
In order to reveal the contribution of the renormalization factor $|Z|$ (via the rescaling coefficient $\Gamma$) to the spectrum, results of the ALDA+LT method are also displayed together with the experimental measurements \cite{Weissker2006,Weissker2010}.
Fig.~\ref{fig:fig4}(a) and (b) show that when the major feature of the spectrum is located below $\sim 32$ eV, the correction from the rescaling coefficient is small.
However, when the spectrum feature goes beyond the energy, the rescaling coefficient contributes a significant increase in the spectrum strength compared to the ALDA+LT results, as shown in Fig.~\ref{fig:fig4}(c) and (d).
It suggests that the improvement mainly comes from the subtle balance between the life-time broadening and the rescaling of renormalization factors.
A collection of DSF of various directions and transferred momenta can be found in the Supplemental Materials \footnote{See Supplemental Materials for DSF for other transferred momenta and directions of Si.}.
In addition, one can also expect that the preceding arguments for the validity regime of approximations also apply to Si. 
As displayed in Fig.~\ref{fig:fig2}, the transition energy is around 32 eV, which is roughly twice of the plasmon frequency of Si about 16.5 eV \cite{kittel_book}. 

\section{Summary}
In summary, we propose a composite ansatz without adjustable parameter to take the full feature of the screened Green's function, especially the renormalization factors, into consideration in the TDDFT calculation framework of DSF. 
We show with Li and Si as illustrating examples that the proposed method provides a much improved  prediction to the DSF for typical metallic and semiconductor systems up to large $|{\bf q}|$.
The new method may bring about interesting possibilities.
For example, it may be used in experiments to directly infer the electronic structure  from the DSF spectrum (with the help of statistical techniques such as the Bayesian inference method \cite{Lindley1972}), by taking the advantage of the relatively simple form of the ansatz and the much improved theoretical results in a broad range of ${\bf q}$.
The new method  also forms an improved theoretical base for the diagnostics of temperature and density inside dense plasmas.

\acknowledgments{
This work is supported by the National Natural Science Foundation of China (NSFC) under Grants No. 12005012, and No. U1930402; and the Laboratory Youth Fund of Institute of Applied Physics and Computational Mathematics under Grant No. 6142A05QN21005.
}

\bibliographystyle{apsrev4-2}
\bibliography{mo}
\end{document}